\documentclass[twoside,reqno]{mbjp}
\usepackage{epsfig,cite}
\usepackage{graphics}
\usepackage{amssymb,amsmath}
\begin{document}

\sloppy \raggedbottom

 \setcounter{page}{1}




\title{Nuclear dependence of the 2p2h electroweak response in the Relativistic Fermi Gas model
\thanks{Contribution presented at the Workshop ``Advanced Aspects in Nuclear Structure and Reactions at Different Energy Scales'', 25-28 April 2017, Arbanasi, Bulgaria}}

\runningheads{M.B. BARBARO {\it et al.}}{NUCLEAR DEPENDENCE OF THE 2P2H RESPONSE}

\begin{start}
\author{M.B. Barbaro}{1,2}, \coauthor{J.E. Amaro}{3}, \coauthor{J.A. Caballero}{4}, \coauthor{A. De Pace}{2}, \coauthor{T.W. Donnelly}{5}, \coauthor{G.D. Meg\'ias}{4}, \coauthor{I. Ruiz Simo}{3}
\address{Dipartimento di Fisica, Universit\`a di Torino, Italy}{1}
\address{Istituto Nazionale di Fisica Nucleare, Sezione di Torino, Italy}{2}
\address{Departamento de Fisica Atomica, Molecular y Nuclear and Instituto de Fisica Teorica y Computacional Carlos I, Universidad de Granada, Spain}{3}
\address{Departamento de Fisica Atomica, Molecular y Nuclear, Universidad de Sevilla, Spain}{4}
\address{Center for Theoretical Physics, Laboratory for Nuclear Science and Department of Physics, Massachusetts Institute of Technology, Cambridge, USA}{5}

\begin{Abstract}
We present the results of a recent study~\cite{Amaro:2017eah} of meson-exchange two-body currents in lepton-nucleus inclusive scattering
at various kinematics and for different nuclei within the Relativistic Fermi Gas model. 
We show that the associated nuclear response functions at their peaks scale as $A k_F^2$, for Fermi momentum $k_F$ going from 200 to  300 MeV/c and momentum transfer $q$ from $2k_F$ to 2 GeV/c. This behavior is different from what is found for the quasielastic response, which scales as $A/k_F$. This result can be valuable in the analyses of long-baseline neutrino oscillation experiments, which need to implement these nuclear effects in Monte Carlo simulations for different kinematics and nuclear targets. 
\end{Abstract}


\end{start}


The study of two-particle two-hole (2p-2h) excitations in lepton-nucleus scattering has gathered renewed attention over the last few years owing to its importance for neutrino oscillation experiments.
While the main purposes of these experiments are the precise measurement of neutrino properties and the exploration of new physics beyond the Standard Model of particle physics, the data analysis strongly relies on the input from nuclear physics.
The analysis of the currently operating (T2K, NOvA) and future (T2HK, DUNE) long-baseline neutrino oscillation experiments requires indeed a very precise knowledge of neutrino-nucleus cross sections over an energy regime going from hundreds to thousands of MeV.
The main difficulty in the interpretation of the data arises from the ignorance of the exact incoming neutrino energy, which is widely distributed around its average value according to a certain flux $\Phi(E_\nu)$. The reconstruction of the neutrino energy from the observed reaction products is obviously strongly dependent upon the nuclear model used in the analysis.
Hence the need for an accurate description of the many-body nuclear system and of different reaction mechanisms, from the quasielastic (QE) region up to the deep-inelastic scattering one. This has motivated an intense recent activity on this subject.
An updated review of the status and challenges of the field can be found in Ref.~\cite{WP}.

2p2h excited states have been extensively explored in the past
\cite{Donnelly:1978xa}-\cite{DePace04}  in electron scattering studies: they correspond to the ejection of two nucleons
above the Fermi level, with the associated creation of two holes inside the Fermi sea, and give a large contribution to the inclusive $(e,e')$ cross
section in the so-called ``dip region'' lying between the QE and $\Delta(1232)$ excitation peaks (see \cite{Meg16e} for an exhaustive comparison with electron scattering data on $^{12}C$).
In neutrino scattering, 
2p-2h excitations have been shown \cite{Mar09}-\cite{Ama10a} to play a crucial role in explaining the cross sections measured in the MiniBooNE, MINERvA and T2K experiments
\cite{Mini1,Mini2,Minerva1,Minerva2,T2K1,T2K2}.

Whereas most of the existing work on the 2p2h contribution to neutrino-nucleus cross section refers to a carbon target
\cite{Mar09,Ama10b,Nie11,Ben15,Roc16,Sim16,Kat16,Meg16e,Meg16nu,Ivanov:2015aya,Ivanov:2015wpa,Ivanov:2016vul,Ivanov:2016mzo},
 it is becoming more and more important to extend the calculations
 to heavier nuclei, in particular argon and oxygen, which also are and will be used as targets in neutrino oscillation experiments.
 The exact evaluation of the 2p2h cross section involves a 7-dimensional integral for each value of the energy and momentum transfer, and an additional integral over the experimental neutrino flux should be performed before comparing with the data. 
 Although this can be done in principle for different nuclear targets, it is useful to provide an estimate of the density dependence of these contributions, which can be used to extrapolate the results from one nucleus to another. This is the main motivation of the present study.

The lepton-nucleus inclusive cross section can be described in terms
of response functions, which embody the nuclear dynamics. 
There are  two response functions in the case of electron scattering,
$R^L$ and $R^T$,
and five in the case of charged-current (anti)neutrino scattering,
$R^{CC}$, $R^{CL}$, $R^{LL}$, $R^{T}$, and $R^{T'}$,
all of them depending upon the momentum and energy transfer $(q,\omega)$.
Each response function is related to specific components of the hadronic tensor $W^{\mu\nu}(q,\omega)$ and receives contributions from different reaction mechanisms (one-body knockout, two-body knockout, resonance excitation, etc.) depending on the kinematics.


Before examining the two-body contribution, it is worth reminding how the one-body cross section depends on the nuclear density.
In \cite{Don99-1,Don99-2} inclusive electron scattering data
from various nuclei were analyzed in terms of ``superscaling'': it
was shown that, for energy loss below the quasielastic peak, the scaling
functions, represented versus an appropriate dimensionless scaling
variable, are not only independent of the momentum transfer (scaling
of first kind), but they also coincide for mass number $A\geq$4 (scaling
of second kind).  More specifically, the reduced QE cross section
was found to scale as
$A/k_F$, $k_F$ being the Fermi momentum. 
It was also shown that for higher energy transfers superscaling is
broken and that its violations reside in the transverse channel rather
than in the longitudinal one. Such violations must be ascribed to
reaction mechanisms different from one-nucleon
knockout. Two-particle-two-hole excitations, which are mainly transverse and
occur in the region between the quasielastic and $\Delta$ production
peaks, are -- at least in part -- responsible for this violation.

\begin{figure}[!htb]
\begin{center}
\epsfig{file=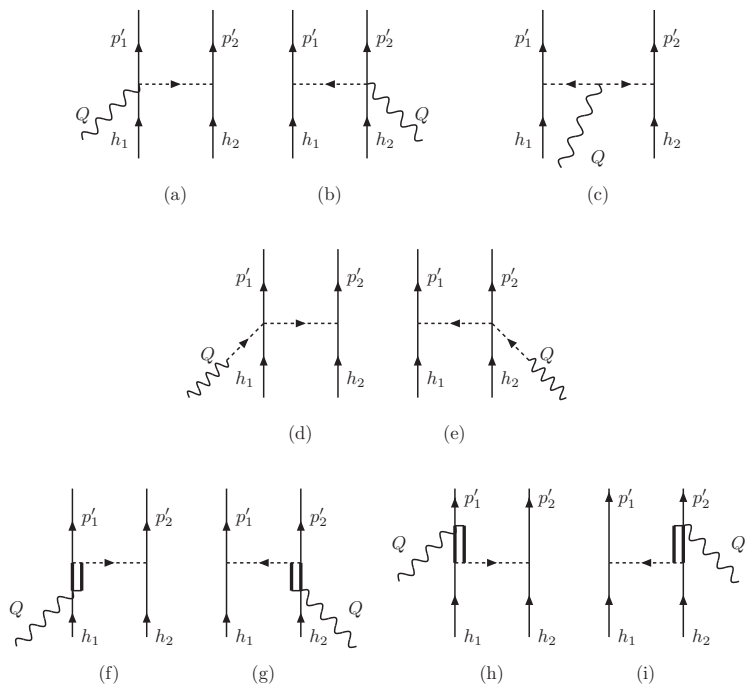,scale=0.7}
\caption{Feynman diagrams for the electroweak MEC model corresponding to neutrino scattering}
\label{fig:fig1}
\end{center}
\end{figure}%

The model we use to evaluate the 2p-2h nuclear responses is based on the Relativistic Fermi Gas (RFG), where it is possible to perform an exact relativistic calculation. It should be stressed that in the GeV energy regime we are interested in these effects cannot be ignored. In the RFG model relativity affects not only the kinematics, but also the nuclear current matrix elements, which are different from the non-relativistic ones. 

The two-body meson exchange currents (MEC) used in this work are represented in Fig.~\ref{fig:fig1} for the weak case (where the wavy line represents a W-boson) and are deduced from a fully relativistic Lagrangian including nucleons (solid lines), pions (dashed lines) and $\Delta$  (thick lines in diagrams f-i) degrees of freedom.
In the electromagnetic case the diagrams (d) and (e) are absent and the wavy line represents the exchanged photon.
These elementary diagrams give rise to a huge number of many-body diagrams,
each of them involving a 7-dimensional integral. In order to speed up the calculation we have recently proposed and tested approximate schemes, capable of reducing the dimensionality of the integral  to 1~\cite{RuizSimo:2017onb} or 3~\cite{MCA}. However here we employ the exact results.
Further details of the model can be found in \cite{DePace03} for the electromagnetic case and in \cite{Sim16} for the extension to the weak sector.

\begin{figure}[!htb]
\begin{center}
\epsfig{file=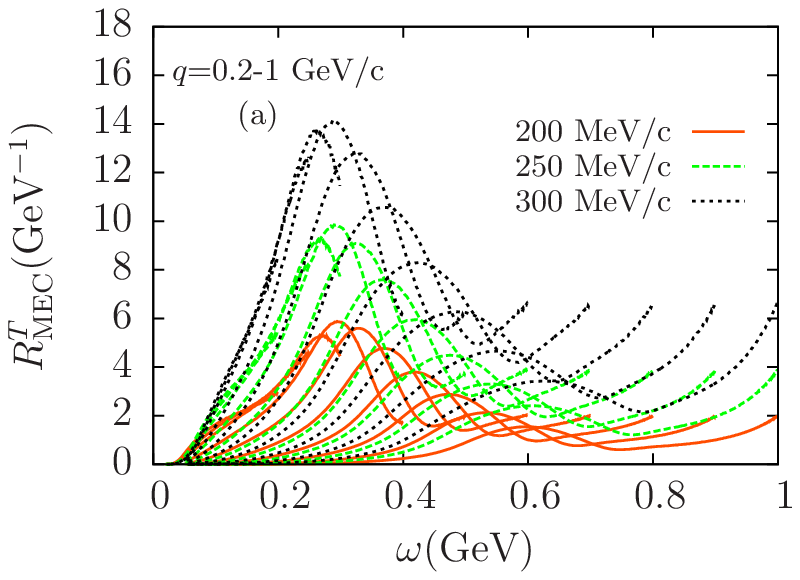,scale=0.7}
\epsfig{file=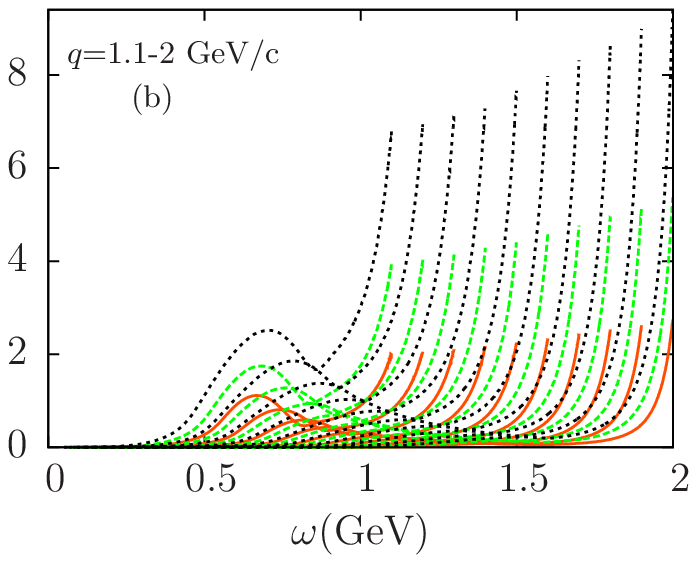,scale=0.7}
\caption{The 2p-2h MEC response plotted versus  $\omega$ for three values of the Fermi momentum $k_F$ and for different values of the momentum transfer $q =$ 200, \ldots, 1000 MeV/c (left panel) and 1100, \ldots 2000 MeV/c (right panel), increasing from left to right.}
\label{fig:fig2}
\end{center}
\end{figure}

Since the behavior with density of the nuclear response is not
expected to depend very much on the specific channel or on the nature
of the probe, for sake of illustration we focus on the electromagnetic
2p-2h transverse response, which largely dominates over the
longitudinal one. 
  Our starting point is therefore the electromagnetic transverse response, $R^T_{\rm MEC}$, which is displayed in Fig.~\ref{fig:fig2} as a function of the
energy transfer $\omega$ for momentum transfers $q$ ranging from 50 to
2000 MeV/c and three values of $k_F$
from 200 to 300 MeV/c, the typical range of Fermi momenta to which most nuclei belong~\cite{Chiara}.

%

\begin{figure}[!htb]
\begin{center}
\epsfig{file=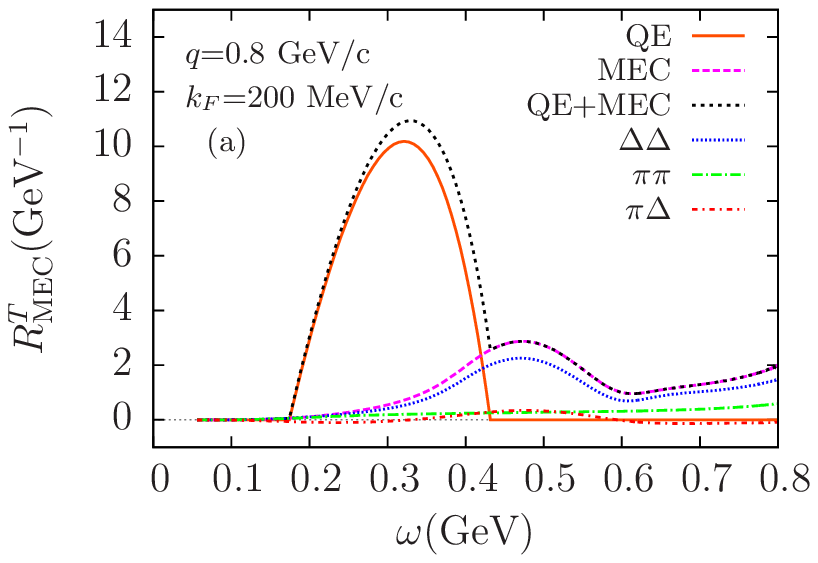,scale=0.7}
\epsfig{file=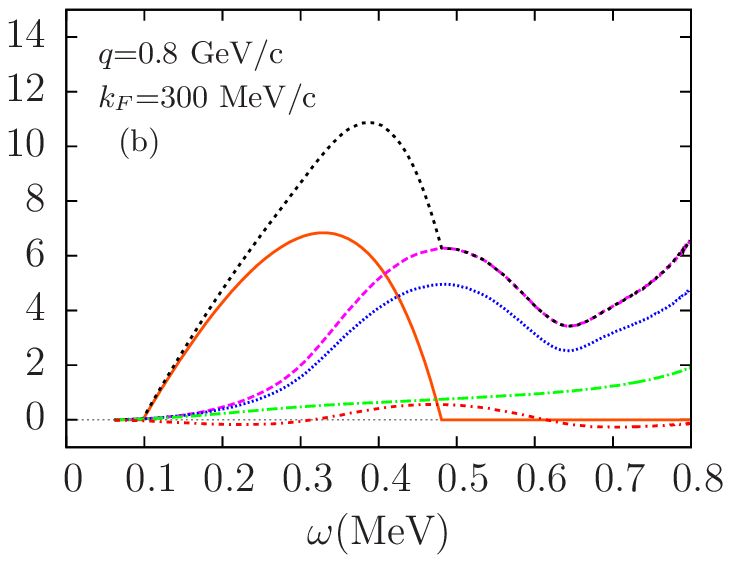,scale=0.7}
\\
\epsfig{file=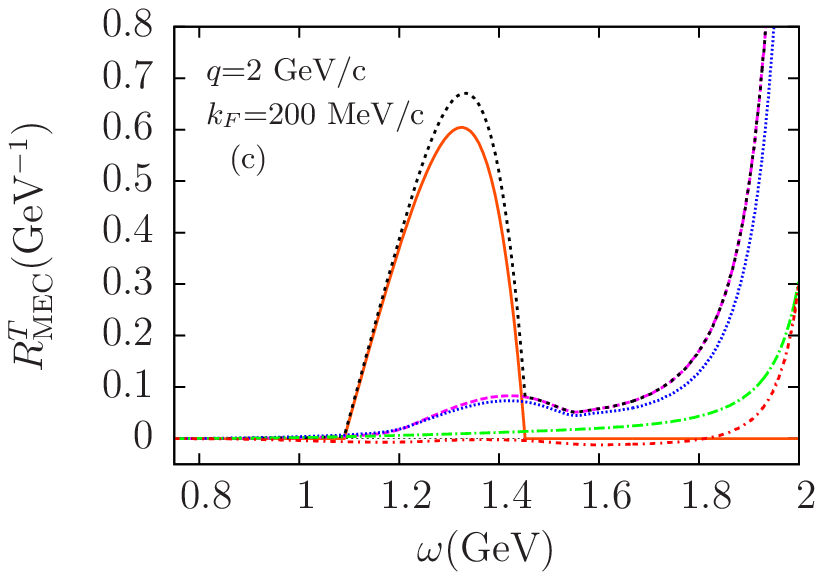,scale=0.7}
\epsfig{file=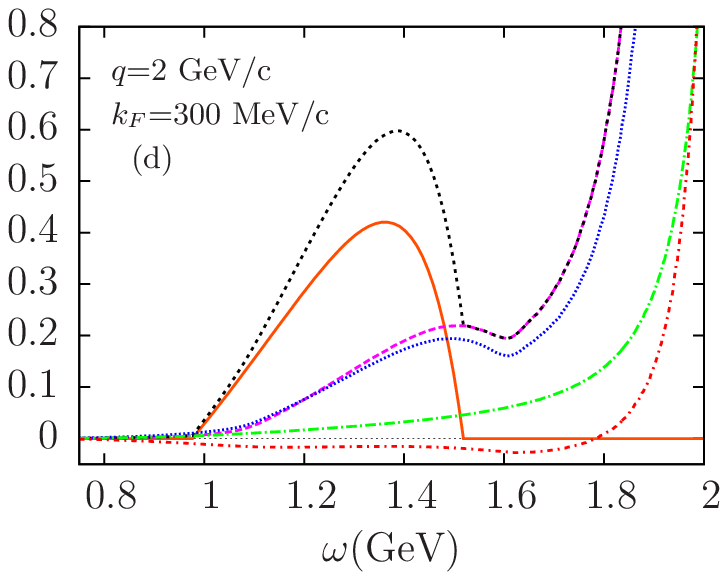,scale=0.7}
\caption{The 2p-2h MEC transverse response $R^T_{\rm MEC}$ and the separate $\Delta\Delta$, $\pi\pi$ and $\pi\Delta$-interference components
plotted versus $\omega$. The free RFG transverse response (red curves) is also shown for reference.}
\label{fig:fig3}
\end{center}
\end{figure}

In order to appreciate the relevance of the MEC response as compared to QE one,
in Fig.~\ref{fig:fig3} we show both response functions for two values of the momentum transfer and two different $k_F$. The MEC response is also splitted into its 
$\Delta\Delta$, $\pi\pi$ and $\pi\Delta$-interference components, showing that the $\Delta\Delta$ contribution is dominant in all cases. It appears that, while for a relatively light nucleus ($k_F$=200 MeV/c) the MEC peak drops from $\sim$30\% to $\sim$10\% of the QE one when the momentum transfer $q$ goes from 0.8 to 2 GeV/c,
for a heavy nucleus ($k_F$=300 MeV/c) the MEC peak is almost as high as the QE one at $q$=0.8 GeV/c and about half of it for $q$=2 GeV/c. This shows that the $k_F$-behavior of the 2p2h response is very different from the above mentioned $1/k_F$ scaling law typical of the QE response (scaling of second kind).
The results in Fig.~\ref{fig:fig3} also show that the 2p2h response drops faster than the quasielastic one as the momentum transfer increases, according to the fact that the MEC also break scaling of first kind.

\begin{figure}[!htb]
\begin{center}
\epsfig{file=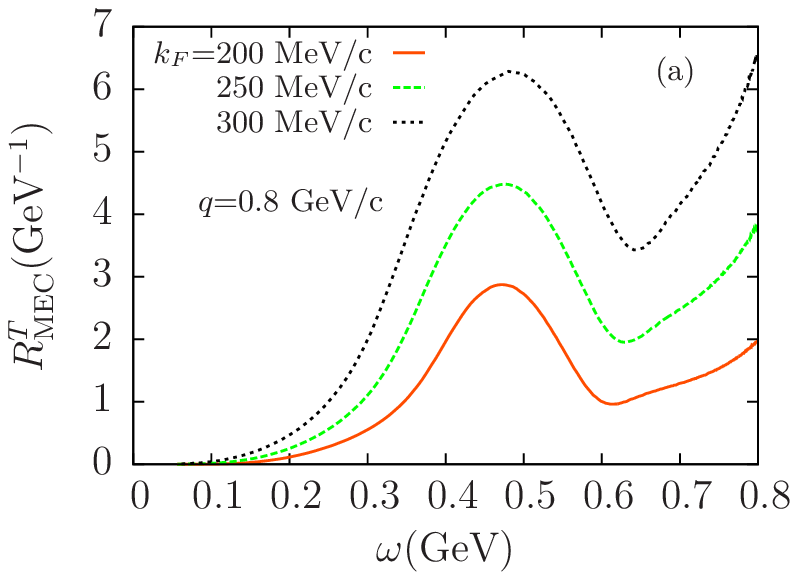,scale=0.7}
\epsfig{file=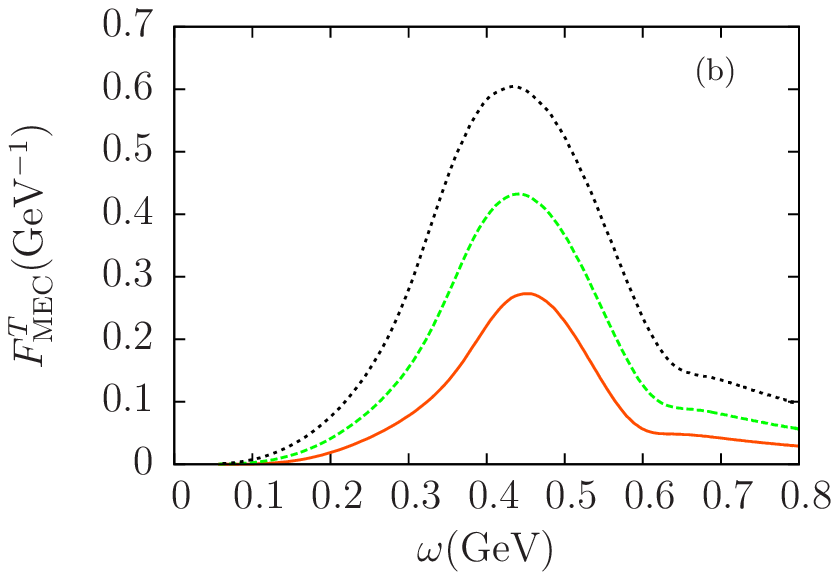,scale=0.7}
\\
\epsfig{file=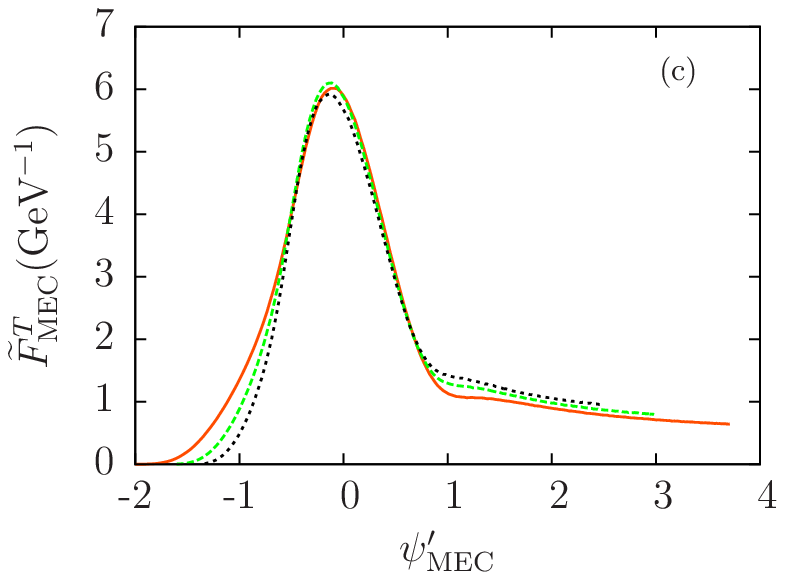,scale=0.7}
\epsfig{file=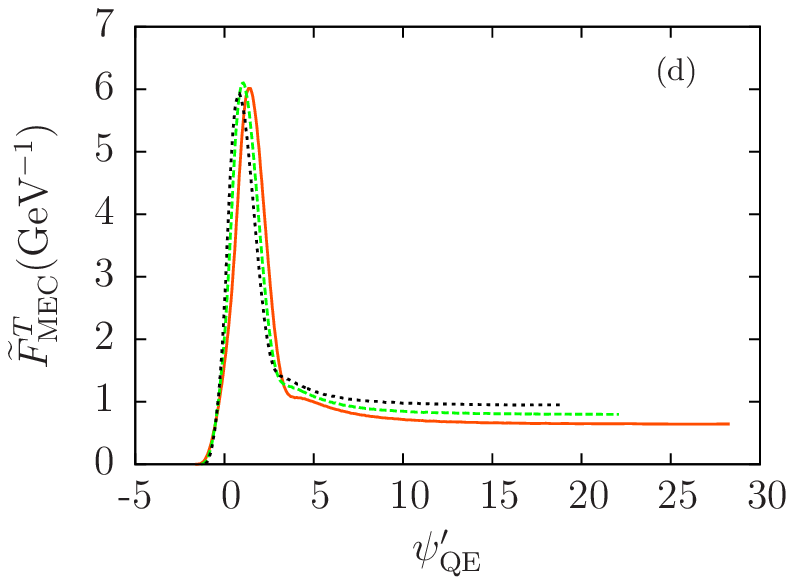,scale=0.7}
\caption{Upper panels: the 2p-2h MEC response (a) and the reduced
   response defined by Eq. (\ref{eq:FT}) (b) plotted versus $\omega$
   for $q$=800 MeV/c and Fermi momentum $k_F$ varying between 200
   (lower curve) and 300 (upper curve) MeV/c.  Lower panels: the
   corresponding scaled 2p-2h MEC response defined by
   Eq. (\ref{eq:tilf}) plotted versus the scaling variables
   $\psi^\prime_{\rm MEC}$ (c) and $\psi^\prime_{\rm QE}$
   (d). }
\label{fig:fig4}
\end{center}
\end{figure}

In order to explore the $k_F$-behavior of $R^T_{\rm MEC}$, we first remove the single-nucleon physics from the problem (which also causes the fast growth of the response as $\omega$ approaches the light-cone) and we define the following reduced response
(per nucleon) 
\begin{equation}
F^T_{\rm MEC}(q,\omega) \equiv
\frac{R^T_{\rm MEC}(q,\omega)}{ Z G_{Mp}^2(\tau) + N G_{Mn}^2(\tau)} \,,
\label{eq:FT}
\end{equation} where $\tau\equiv (q^2-\omega^2)/(4m_N^2)$ and
$G_{Mp}$ and $G_{Mn}$ are the proton and neutron magnetic form
factors. For simplicity here we neglect in the single-nucleon dividing 
factor small contributions coming from the motion of the nucleons,
  where the electric form factor contributes, which depend on
  the Fermi momentum \cite{scaling}.
In the upper panels of Fig.~\ref{fig:fig4} we
show the response $R^T_{\rm MEC}$ and the reduced response $F^T_{\rm
  MEC}$ for $q$=800 MeV/c and the same three values of $k_F$ used above. It clearly
appears that the 2p-2h response, unlike the 1-body quasielastic one,
increases as the Fermi momentum increases.   In the lower panels of Fig.~\ref{fig:fig4} we display the scaled 2p-2h
MEC response, defined as
\begin{equation} \widetilde F^T_{\rm
  MEC}\left(\psi^\prime_{\rm MEC}\right) \equiv \frac{F^T_{\rm
    MEC}}{\eta_F^2} \,,
\label{eq:tilf}
\end{equation} namely the reduced response divided by $\eta_F^2 \equiv
(k_F/m_N)^2$, as a function of the MEC scaling variable
$\psi^\prime_{\rm MEC}(q,\omega,k_F)$ (left panel) and of the
quasielatic one $\psi^\prime_{\rm QE}(q,\omega,k_F)$ (right panel).
The MEC scaling variable is defined in Ref.~\cite{Amaro:2017eah}, in analogy with the usual QE scaling variable \cite{scaling}.
  The results
show that the reduced 2p-2h response per nucleon roughly scales as $k_F^2$ when
represented as a function of $\psi^\prime_{\rm MEC}$ (Fig.~\ref{fig:fig4}c), {\it i.e.}, the scaled 2p-2h MEC response shown there coalesces at the peak into a universal result. This
scaling law is very accurate at the peak of the 2p-2h response, while
it is violated to some extent at large negative values of the scaling
variable. Fig.~\ref{fig:fig4}d shows that in this ``deep scaling'' region it is more appropriate to
use the usual scaling variable $\psi^\prime_{\rm QE}$ devised for
quasielastic scattering.
  This latter region was previously investigated in \cite{DePace04}, where the specific cases of $^{12}$C and $^{197}$Au were considered and the results were compared with 
JLab data at electron energy $\epsilon$=4.045 GeV: there it was shown that
at very high momentum transfers the 2p-2h MEC contributions are very significant in this deep scaling region, to the extent that they may even provide the dominant effect. Nevertheless, the scaling violations associated to them were shown to be reasonably compatible with the spread found in the data.

A closer inspection of the scaling properties of the 2p-2h response, performed in Ref.~\cite{Amaro:2017eah}, has also shown that all the contributions ($\Delta\Delta$, $\pi\pi$ and $\pi\Delta$) roughly grow as $k_F^2$, the quality of scaling being better for the $\Delta\Delta$ piece than for the other two
contributions. Furthermore, at high momentum
transfer the total MEC response scales better than the pure $\Delta$
piece around the peak, indicating a compensation of scaling violations
between the three terms.

\begin{figure}[!htb]
\begin{center}
\epsfig{file=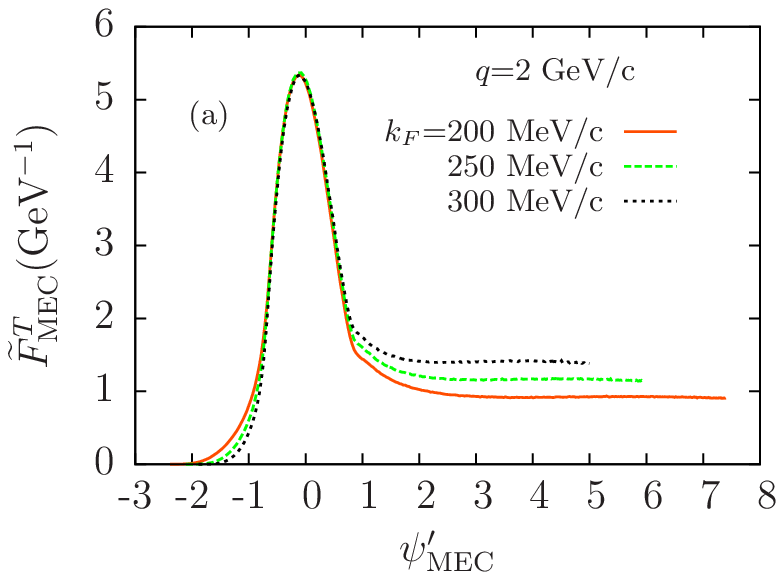,scale=0.7}
\epsfig{file=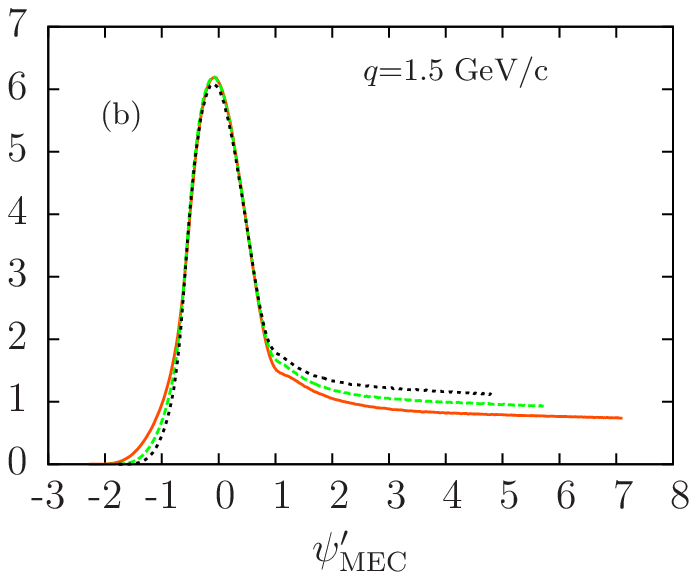,scale=0.7}
\\
\epsfig{file=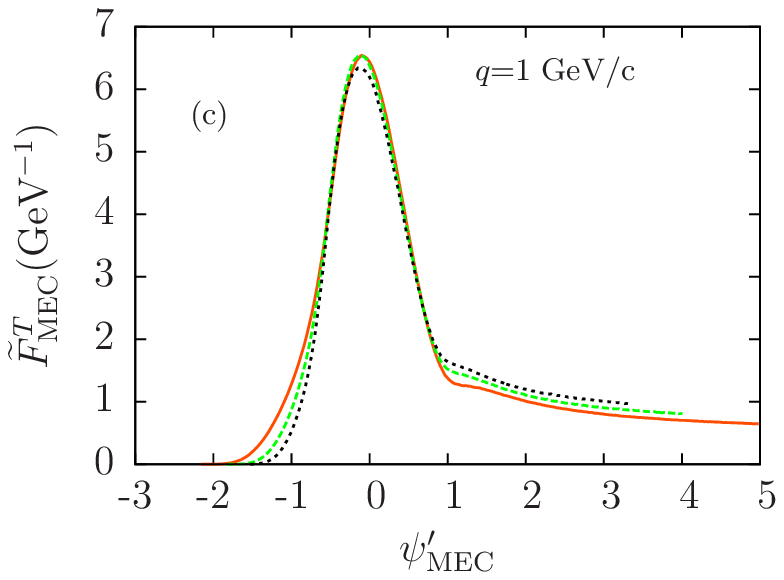,scale=0.7}
\epsfig{file=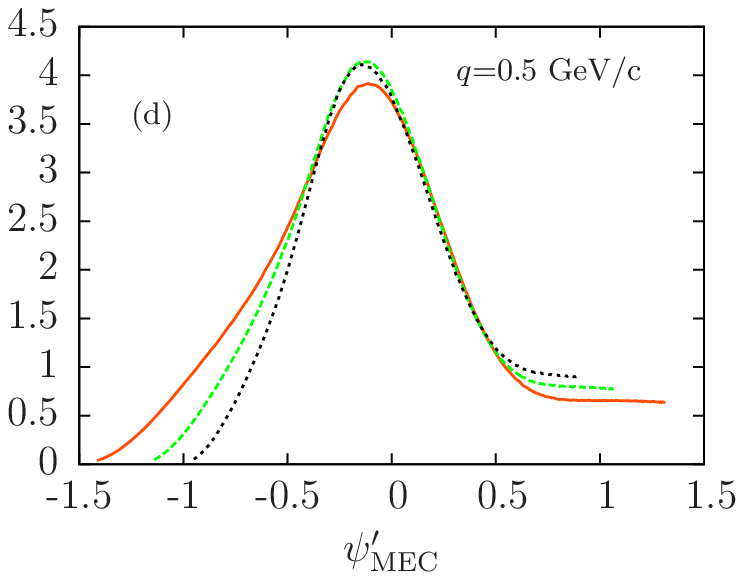,scale=0.7}
\caption{As for Fig.~\ref{fig:fig4}c, but now for different values of $q$.}
\label{fig:fig5}
\end{center}
\end{figure}

In Fig.~\ref{fig:fig5} the scaled 2p-2h MEC response is now plotted versus $\psi^\prime_{\rm MEC}$ for four values of $q$. Here we see 
that the same $k_F$-dependence is valid for different values of $q$ as long as Pauli blocking is not active, namely $q>2 k_F$. At lower $q$ and in the deep scaling region this type of scaling is seen to be broken.

\begin{figure}[!htb]
  \begin{center}
\epsfig{file=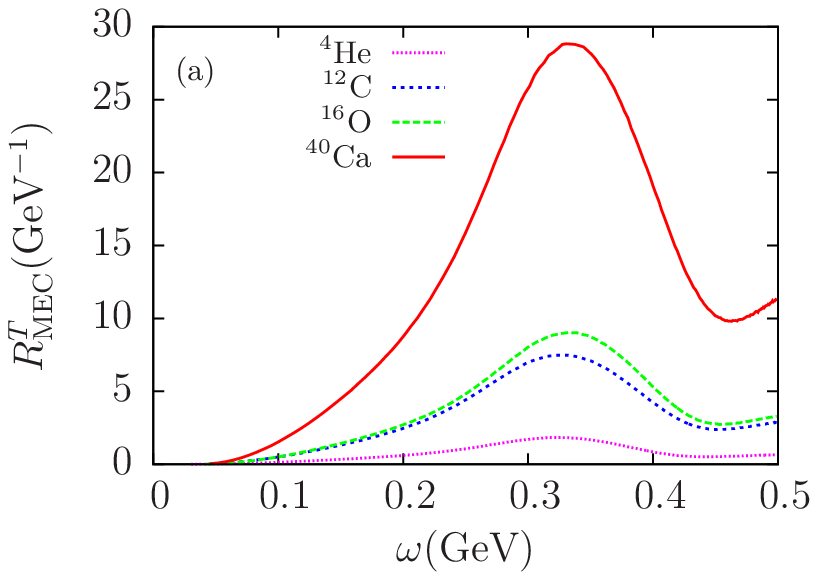,scale=0.4}
\epsfig{file=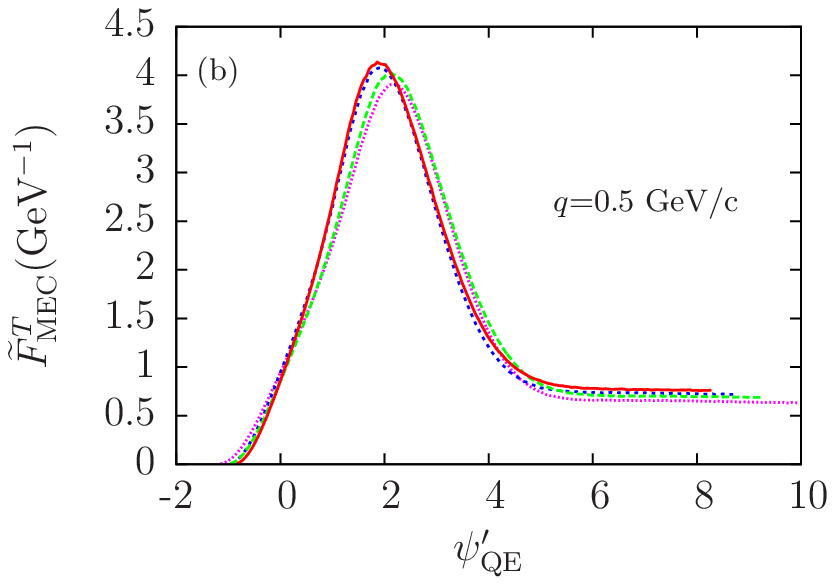,scale=0.4}
\epsfig{file=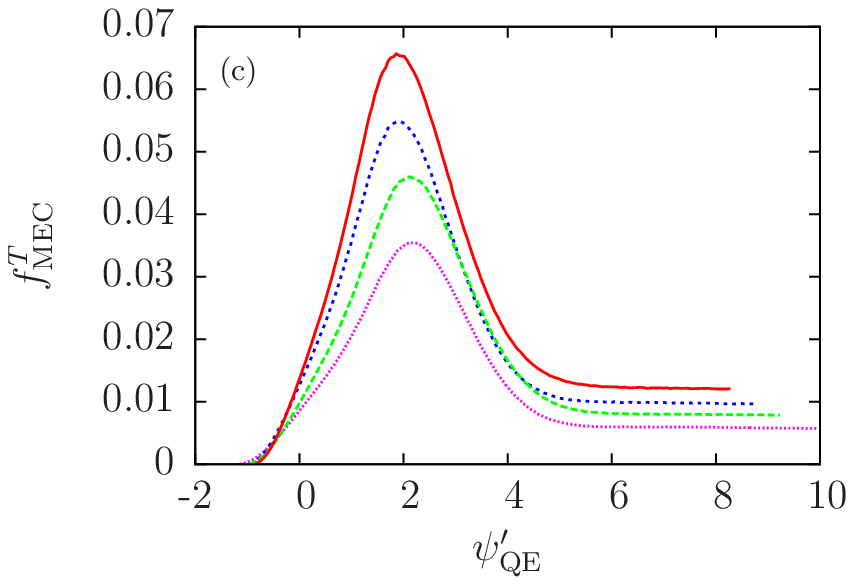,scale=0.4}
\\
\epsfig{file=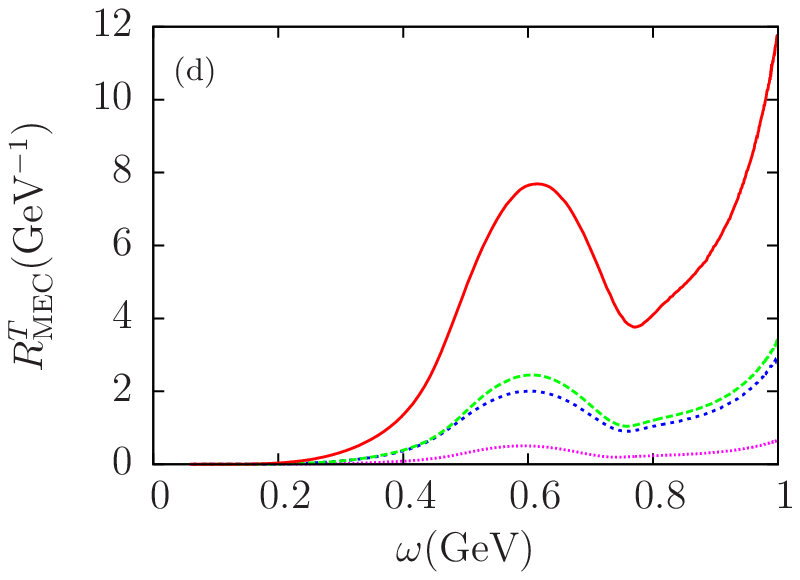,scale=0.4}\hspace{0.2cm}%
\epsfig{file=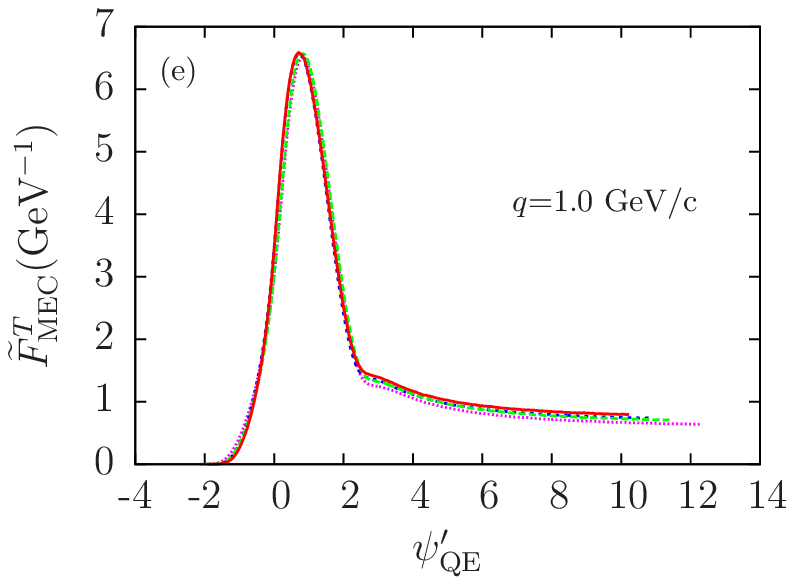,scale=0.4}\hspace{0.1cm}%
\epsfig{file=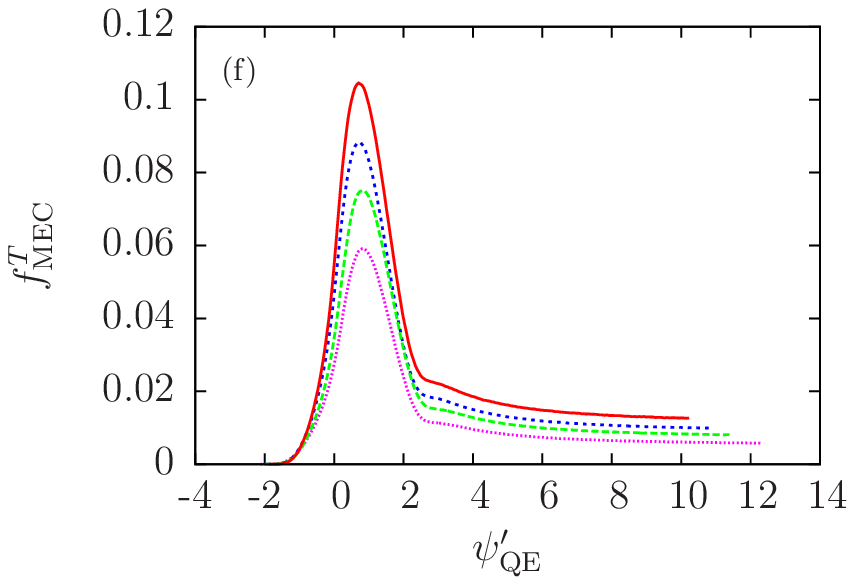,scale=0.4}
\\
\epsfig{file=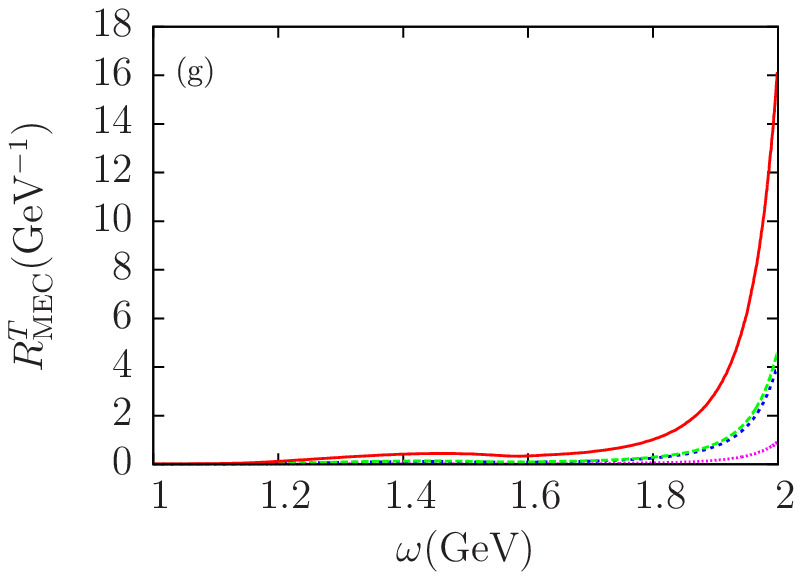,scale=0.4}\hspace{0.1cm}%
\epsfig{file=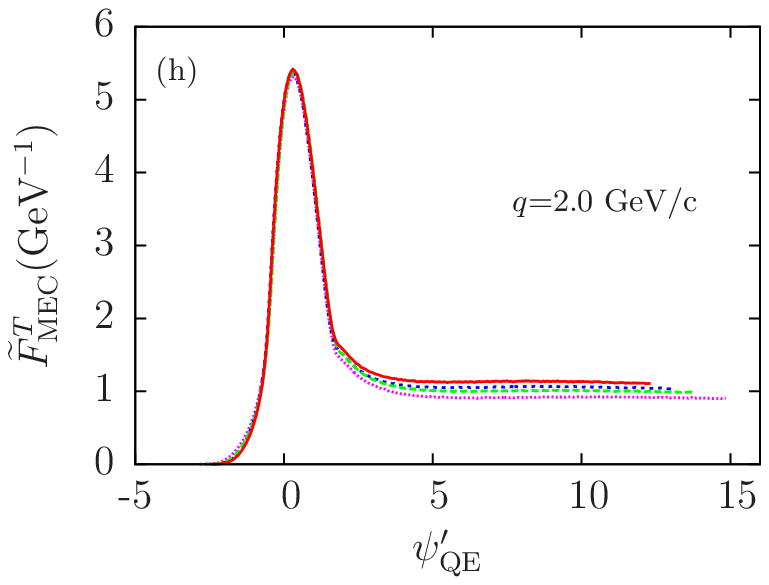,scale=0.4}\hspace{0.2cm}%
\epsfig{file=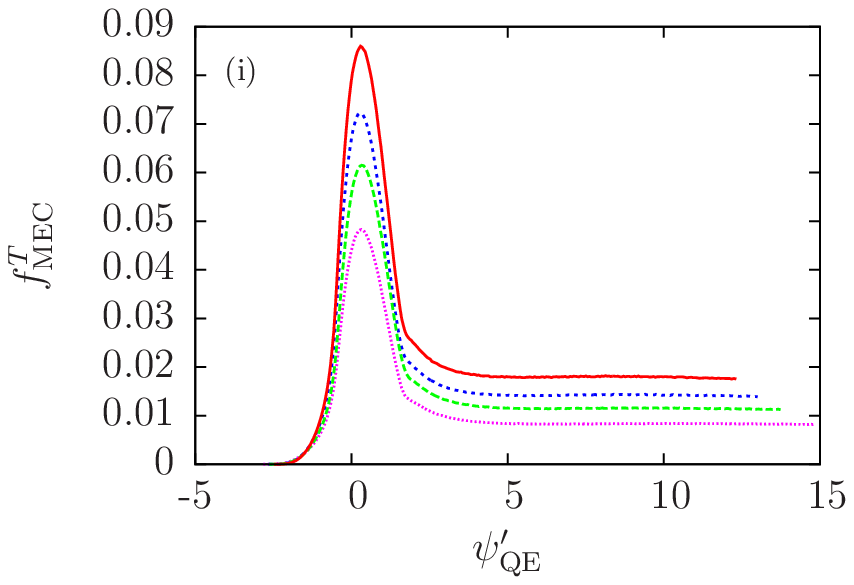,scale=0.4}
    \caption{The 2p-2h MEC response (first column), the corresponding scaled response $\widetilde F^T_{\rm MEC}$ defined by Eq. (\ref{eq:tilf}) (second column) and the superscaling function $f^T_{MEC}$ defined by Eq. (\ref{eq:fsup})
    (third column) for four nuclei and three values of momentum transfer $q$.}
\label{fig:fig6}
\end{center}
\end{figure}

Finally, focusing on practical cases, in Fig.~\ref{fig:fig6} we show $R^T_{\rm MEC}$ versus $\omega$, together with $\widetilde F^T_{\rm MEC}$ and
\begin{equation}
f^T_{MEC}\equiv F^T_{MEC}\times k_F
\label{eq:fsup}
\end{equation}
versus $\psi^\prime_{\rm QE}$ for three values of $q$ and for the symmetric nuclei $^{4}$He, $^{12}$C, $^{16}$O and $^{40}$Ca.
The case of asymmetric nuclei, $Z\neq N$ requires more involved
formalism and will be addressed in future work, although preliminary
studies indicate that the qualitative behavior with $k_F$ does not
change dramatically unless $N-Z$ is very large. 
The cases of $^{12}$C and $^{16}$O are clearly relevant for ongoing neutrino oscillation studies, whereas the case of $^{40}$Ca is a symmetric nucleus lying close to the important case of $^{40}$Ar.
For comparison, $^{4}$He is also displayed and, despite its small mass, is seen to be ``typical''. In contrast, the case of $^{2}$H, whose Fermi momentum is unusually small ($k_F$= 55 MeV/c), was also explored and found to be completely anomalous: the MEC responses ($R^T_{\rm MEC}$) and superscaling results ($f^T_{\rm MEC}$) were both too small to show in the figure.

%


Summarizing, the 2p-2h MEC response
function per nucleon roughly grows as $k_F^2$ for Fermi momenta
varying from 200 to 300 MeV/c.  This scaling law is excellent around
the MEC peak for high values of $q$, it starts to break down around $q = 2
k_F$, and gets worse and worse as $q$ decreases.  This behavior must
be compared with that of the 1-body response, which scales as $1/k_F$:
hence the relative importance of the 2p-2h contribution grows as
$k_F^3$.  This result allows one to get an estimate of the relevance
of these contributions for a variety of nuclei, of interest in ongoing
and future neutrino scattering experiments, and should facilitate the
implementation of 2p-2h effects in event generators.

\section*{Acknowledgements}

This work was partially supported by the INFN under Project MANYBODY, by the University of Turin under Project BARM-RIC-LOC-15-02, by the Spanish Ministerio de Economia y Competitividad and ERDF (European Regional Development Fund) under contracts FIS2014-59386-P, FIS2014-53448-C2-1, by the Junta de Andalucia (grants No. FQM-225, FQM160), and part (TWD) by the U.S. Department of Energy under cooperative agreement DE-FC02-94ER40818. IRS acknowledges support from a Juan de la Cierva-incorporacion program from Spanish MINECO. GDM acknowledges support from a Junta de Andalucia program (FQM7632, Proyectos de Excelencia 2011).

\end{document}